\documentclass[prl,aps,groupedaddress,showpacs,preprintnumbers,twocolumn,%
nofootinbib,superscriptaddress]{revtex4}%
\usepackage{graphicx}
\usepackage{amsmath}
\usepackage{bm}
\usepackage{epsfig}
\usepackage{amssymb}

\begin{document}
\title{Factorization theorems for exclusive heavy-quarkonium production}
\author{Geoffrey~T.~Bodwin}
\affiliation{High Energy Physics Division, Argonne National Laboratory,\\
9700 South Cass Avenue, Argonne, Illinois 60439, USA}
\author{Xavier \surname{Garcia i Tormo}}
\affiliation{High Energy Physics Division, Argonne National Laboratory,\\
9700 South Cass Avenue, Argonne, Illinois 60439, USA}
\author{Jungil~Lee}
\affiliation{Department of Physics, Korea University, Seoul 136-701, Korea}

\date{\today}

\preprint{ANL-HEP-PR-08-28}

\begin{abstract}
We outline the proofs of the factorization theorems for exclusive
two-body charmonium production in $B$-meson decay and $e^+e^-$
annihilation to all orders in perturbation theory in quantum
chromodynamics. We find that factorized expressions hold up to
corrections of order $m_c/m_b$ in $B$-meson decay and corrections of
order $m_c^2/s$ in $e^+e^-$ annihilation, where $m_c$ is the charm-quark
mass, $m_b$ is the bottom-quark mass, and $\sqrt{s}$ is the $e^+e^-$
center-of-momentum energy.
\end{abstract}

\maketitle

The exclusive decays of $B$ mesons into a light meson plus a charmonium
are significant tools for understanding the weak interactions. For
example, they could provide new constraints on the
Cabibbo-Kobayashi-Maskawa (CKM) matrix and enhance our understanding of
the origins of CP violation.
However, the effects of the strong interactions are significant in such
processes and present an obstacle to achieving reliable theoretical
calculations of the process rates. The exclusive production of
double-charmonium in $e^+e^-$ annihilation provides an
arena in which to explore the mechanisms of charmonium production
and the interplay between the perturbative and nonperturbative regimes
of quantum chromodynamics (QCD). However, measurements of the cross
sections for double-charmonium production by the Belle \cite{Abe:2004ww}
and BABAR \cite{Aubert:2005tj} Collaborations have spurred a
re-examination of the theoretical predictions for those cross sections
\cite{Bodwin:2007ga}.

In theoretical computations of the rates for the exclusive decays of $B$
mesons into a light meson plus a charmonium and of the rates for
the exclusive  production of double-charmonium in $e^+e^-$
annihilation, a crucial step is the separation of the effects of the
strong interactions into short-distance, perturbatively calculable
contributions and long-distance, inherently nonperturbative
contributions. Such separations are usually embodied in factorization
theorems. In the case of the exclusive decays of $B$ mesons into a light
meson plus a charmonium, several factorization theorems have been
posited \cite{Beneke:2000ry,Chay:2000xn,Bobeth:2007sh}. In the case of
the exclusive  production of double-charmonium in $e^+e^-$
annihilation, factorization conjectures have generally been formulated
in terms of nonrelativistic QCD (NRQCD) \cite{Bodwin:2007ga}. In this
letter, we outline the proofs of the factorization theorems for these
processes in QCD to all orders in perturbation theory. 

We wish to show that the amplitudes for these processes can be written
in a factorized form. The meaning of this statement, for the case of
exclusive decays of $B$ mesons into a light meson plus a charmonium,
is that the decay amplitude is decomposed into the sum of
products of a $B$-meson-to-light-meson form factor and an amplitude for a
charm-quark-antiquark ($c\bar c$) pair to be produced at short-distances
in a color-singlet state and evolve into a charmonium. The form
factor contains a term that can be decomposed into a convolution of a 
hard-scattering amplitude with $B$-meson and light-meson light-cone amplitudes 
and a term that cannot be decomposed further. (These two terms are 
analogous to the two terms in the factorization formula in Eq.~(4) of
Ref.~\cite{Beneke:2000ry} for the case of decays to two light mesons.)  
The $c\bar c$-to-charmonium amplitudes can
be further decomposed into a sum of products of long-distance NRQCD
matrix elements times short-distance coefficients. We argue that this
factorized form holds up to corrections of relative order $m_c/\sqrt{s}$,
where $s=m_b^2$. This result was suggested previously in
Ref.~\cite{Beneke:2000ry}. For the case of the exclusive production of
double-charmonium in $e^+e^-$ annihilation, factorization means
that the production amplitude decomposes into a sum of products of a
short-distance amplitude and two amplitudes for a $c\bar c$ pair to be
produced at short-distances in a color-singlet state and evolve into a
charmonium. Again, the $c\bar c$-to-charmonium amplitudes can be
further decomposed into a sum of products of long-distance NRQCD matrix
elements times short-distance coefficients. We argue that this
factorized form holds up to corrections of relative order $m_c^2/s$,
where $\sqrt{s}$ is the center-of-momentum (CM) energy of the $e^+e^-$
pair. Although our analyses are for the specific cases of $B$ decay and
$e^+e^-$ annihilation, the techniques that we describe should apply to
other exclusive quarkonium production processes and may also shed 
light on factorization in inclusive  quarkonium production. We note that, 
because we consider exclusive two-body decays, we avoid the issues raised in
Ref.~\cite{Nayak:2005rt} concerning light particles co-moving with the
charmonium and the issues raised in Ref.~\cite{Nayak:2007mb} concerning
an additional heavy quark co-moving with the charmonium.

We carry out our analyses in the $B$-meson rest frame and in the CM
frame of the $e^+e^-$ pair, choosing the three-momentum of the light
meson or one of the charmonia to be in the negative $3$ direction and
choosing the three-momentum of the other charmonium to be in the
positive $3$ direction. For a momentum $k$, we define light-cone
momentum components $k^{\pm}=(k^0\pm k^3)/\!\sqrt{2}$, $\bm{k}_\perp$. 
We model the $B$ meson as an on-shell active bottom quark, which
participates in the electroweak interaction, and an on-shell
spectator light antiquark, which does not participate in the
electroweak interaction. We take the quark and antiquark to be in a
color-singlet state. We take the bottom quark to have momentum $p_b$,
with $p_b^0=m_b$ and all other components of momentum zero. We take the
spectator quark to have momentum $p_l$, with
$p_{l\,\mu}\!\sim\!\Lambda_{\rm QCD}$, the QCD scale. We neglect
$\Lambda_{\rm QCD}$ and the difference between $m_b$ and the $B$-meson
mass in comparison with $m_b$. Similarly, we model the light meson as an
on-shell active light quark and an on-shell spectator light antiquark,
with the quark and antiquark in a color-singlet state. We take the
active-quark momentum to be $p_{k_1}\!=\!yp_k+q_k$ and the
spectator-quark momentum to be
$p_{k_2}\!\!=\!\!(1\!\!-\!\!y)p_k-q_k$, where $p_k$ is the
light-meson momentum. $p_k^-\!\sim\!m_b$, $q_{k}^-=0$, $p_{k\perp}=0$,
$q_{k\perp}\sim \Lambda_{\rm QCD}$, and
$p_k^+\!\sim\!q_k^+\!\sim\!\Lambda_{\rm QCD}^2/m_b$.
Finally, we model a charmonium as an on-shell $c\bar c$ pair in a
color-singlet state, with the momentum of the $c$ ($\bar{c}$) equal to
$p_i$ ($\bar p_i$). We take $p_i=P_i/2+q_i$ and $\bar{p}_i=P_i/2-q_i$,
where $P_i$ is the charmonium momentum and $P_i\cdot q_i=0$.  In the
charmonium rest frame, $q_i$ has only spatial components, whose
magnitudes are of order $m_cv$, where  $v$ is the typical charm-quark
velocity in the charmonium rest frame ($v^2\approx 0.3$). In the
$e^+e^-$ CM frame or $B$-meson rest frame, $P_1^+\sim P_2^-\sim Q$,
$P_{1\perp}=P_{2\perp}=0$, and $P_1^-\sim P_2^+\sim m_c^2/Q$, where
$Q=\sqrt{s}$ is the large momentum scale. One can think of the on-shell
amplitude that we use in our model as the on-shell perturbative QCD
amplitude that is matched to 
a soft-collinear-effective-theory amplitude
in the case of the light meson, a heavy-quark-effective-theory
amplitude in the case of the $B$ meson, and an NRQCD amplitude in the
case of the charmonia.

The next step is to identify the regions of loop momenta that give
contributions that are leading in powers of $Q$ when we dress the
lowest-order decay and production amplitudes in these models with
additional gluons. These leading momentum regions correspond to
particular Feynman-diagram topologies
\cite{Sterman:1978bi,Sterman:1978bj}, which we describe below.
The identification
of leading regions has been discussed in detail in
Ref.~\cite{Collins:1989gx}. For a loop momentum $k$, the leading regions
are the {\it hard region}, in which $k_\mu\sim Q$; the {\it
collinear-to-plus} ({\it minus}) {\it region}, in which $|k^+|\!\gg\!
|k^-|$ ($|k^-|\!\gg\! |k^+|$), $k^+k^-\!\sim\! \bm{k}_\perp^2\!\ll\!
Q^2$, and the {\it soft region}, in which $k^+\!\sim\! k^-\!\sim\!
\!|\bm{k}_\perp|\! \ll\! Q$. The collinear regions correspond to the
directions of the momenta of the final-state light meson and charmonia.
We note that, in the case of $B$-meson decays, there is also a
``semi-hard region'' in which propagators are off shell by an amount of
order $m_b\Lambda_{\rm QCD}$. We treat this semi-hard region as part of
the hard region. 
The ``Glauber'' region is
also leading in power counting \cite{Bodwin:1981fv,Bodwin:1984hc}. In
this region, $|k^+|\!\ll\! |k_\perp|$, $|k^-|\!\ll\! |k_\perp|$,
$k_\perp^2\!\ll\! Q^2$. However, for the exclusive processes we are
considering here, the contours of integration of loop momenta are not
pinched in the Glauber region, and it is possible to deform them out of
it on a diagram-by-diagram basis. This is in contrast with the situation
in, for instance, the Drell-Yan process, for which such a
diagram-by-diagram contour deformation is not possible
\cite{Bodwin:1984hc,Collins:1985ue,Collins:1989gx}. Therefore, we
ignore the Glauber region in the remainder of our discussion.
Contributions from the hard (semi-hard) region involve propagator
denominators that are of order $s$ ($m_b\Lambda_{\rm QCD}$) and can be
calculated in perturbation theory. Integrations of gluon momenta over
the soft region or a collinear region that is associated with a massless
particle encounter singularities in propagator denominators that result
in logarithmic divergences. [The soft singularities are at $k_\mu=0$, and
the collinear singularities are at $k^-=0$, $\bm{k}_\perp=0$ ($k^+=0$,
$\bm{k}_\perp=0$) for the collinear-to-plus (to-minus) region.] Thus,
contributions from the soft region and the collinear regions that are
associated with a massless particle cannot be computed reliably in
perturbation theory. The essence of the proofs of the factorization
theorems that we consider here is to show that these contributions
cancel or can be absorbed into the $B$-meson-to-light-meson form factor
or the nonperturbative NRQCD matrix elements. When a gluon three-momentum
becomes parallel to the three-momentum of a charm quark or charm
antiquark, the potential collinear singularity is shielded by $m_c$.
Hence, the contributions from such collinear  regions can be computed
reliably in perturbation theory. Nevertheless, it is important for the
arguments that we will make below to treat these regions separately 
from $H$. As we shall see, the contributions from these regions cancel.

\begin{figure}
\epsfig{file=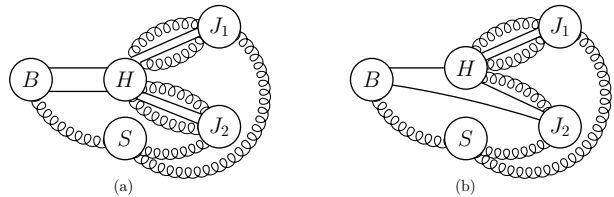,width=8cm}
\vspace*{8pt}
\caption{\label{fig:regions}%
Leading regions for the decay of a $B$ meson to a light meson and a
charmonium.
}
\end{figure}
\begin{figure}
\epsfig{file=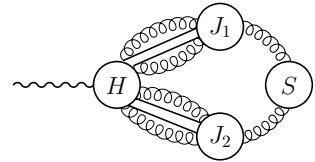,width=4cm}
\vspace*{8pt}
\caption{\label{fig:regionsdc}%
Leading regions for double-charmonium production in $e^+e^-$
annihilation. The wavy line represents the virtual photon.
}
\end{figure}

Now let us specify the diagrammatic topologies that correspond to the
leading regions. We work in the Feynman gauge. There are two
distinct topologies in the case of $B$-meson decays: one in which the
$B$-meson and light-meson spectators participate in the hard interaction
and another in which they do not. These topologies are shown in
Figs.~\ref{fig:regions}(a) and \ref{fig:regions}(b), respectively. 
After one has achieved factorization, the contribution from the first
topology can be decomposed into an expression that contains the
convolution of a hard-scattering amplitude with $B$-meson and
light-meson light-cone amplitudes, while the contribution from the
second topology cannot be decomposed further. The topology for the case
of double-charmonium production is represented in
Fig.~\ref{fig:regionsdc}. The components of the topologies are as 
follows: a $B$
meson ($B$); jet sub-diagrams for each of the collinear regions,
corresponding to a charmonium ($J_1$) and  a light meson ($J_2$) in
Fig.~\ref{fig:regions} and the two charmonia ($J_1$ and $J_2$) in
Fig.~\ref{fig:regionsdc}; a hard sub-diagram ($H$) that includes the
lowest-order annihilation or production process; and a soft sub-diagram
($S$). In $H$, all propagator denominators are of order $s$ or
$m_b\Lambda_{\rm QCD}$. $J_i$ contains the active- and
spectator-quark lines for a given meson or charmonium,  as well as
gluons and loops involving quarks and ghosts with momenta collinear to
the meson or charmonium. $J_i$ attaches to $H$ through the active- and
spectator-quark lines in the topology of Fig.~\ref{fig:regions}(a),
through the active-quark lines in the topology of
Fig.~\ref{fig:regions}(b) and through any number of gluons. $S$ includes
gluons with soft momenta and loops involving quarks and ghost with soft
momenta. $S$ attaches to $J_i$ and to the $B$-meson quark and
antiquark lines through any number of soft gluon lines. Recall that the
$B$-meson spectator antiquark itself carries a soft momentum. In
Fig.~\ref{fig:regions}(b), the light-antiquark-spectator line carries
a momentum with minus component of order $\Lambda_{\rm QCD}$. Such a
momentum can arise either through an endpoint contribution, in which
$p_{k_2}^-\sim \Lambda_{\rm QCD}$ \cite{Beneke:2000ry,Chernyak:1990ag},
or through the soft-collinear-messenger mechanism \cite{Becher:2003qh},
in which an emitted gluon carries away most of $p_{k_2}$.

At this point we can outline the proof of factorization. Suppose that a
gluon from $J_i$ enters $H$. Then we can apply a collinear
approximation to that gluon
\cite{Bodwin:1984hc,Collins:1985ue,Collins:1989gx}. Specifically, we
replace $g_{\mu\nu}$ in the gluon-propagator numerator with $k_\mu \bar
n_{i\nu}/k\cdot \bar n_i$, where $k$ is the gluon momentum, $\bar n_1$
($\bar n_2$) is a unit light-like vector in the minus (plus) direction.
The index $\mu$ corresponds to the attachment of the gluon to $H$, and
the index $\nu$ corresponds to the attachment of the gluon to $J_i$. For
$k$ in the collinear region, the collinear approximation is valid at
leading order in $Q$. Having made the collinear approximation, one can
use the factor $k_\mu$ and the diagrammatic Ward identity (Feynman
identity) for fermion lines $k\cdot\gamma=[(q+k)\cdot
\gamma+m]-[q\cdot\gamma+m]$ and its generalizations to non-Abelian gauge
theories to show that the gluons with collinear momenta decouple from
$H$ \cite{Bodwin:1984hc,Collins:1985ue,Collins:1989gx}. In
general, the decoupled gluons attach to light-like eikonal lines (path 
integrals of the gauge field) that
connect to the point at which an active parton enters $H$. However,
in our case these eikonal lines cancel because the light meson
and the charmonia are color-singlet states. Now suppose that a gluon
from $S$ attaches to a jet sub-diagram at a line with momentum
$p$. We can apply a soft approximation to that gluon
\cite{Collins:1981uk}. Specifically, we replace $g_{\mu\nu}$ in the
gluon-propagator numerator with $k_\mu p_\nu/k\cdot p$, where $k$ is the
gluon momentum, the index $\mu$ corresponds to the attachment of the
gluon to $J_i$ and the index $\nu$ corresponds to the attachment of the
gluon to $S$. We would like to use the factor $k_\mu$ in the soft
approximation and the diagrammatic Ward identities to decouple the soft
gluons.  However, in general, the momentum $p$ is different for every
line to which the soft gluon attaches. That is, the soft approximation,
unlike the collinear approximation, is not independent of the line to
which the $\mu$  end of the gluon attaches. Nevertheless, we can
apply the same soft approximation to all lines in the
collinear-to-light-meson jet, since they are all proportional to the
same light-like vector (up to corrections of order $\Lambda_{\rm
QCD}/m_b$). For the constituents of each charmonium, we can also apply
the same soft approximation up to corrections of order $m_c/Q$. The
reason for this is that, in going from the rest frame of the charmonium
to the $e^+e^-$ CM frame or $B$-meson rest frame, the momenta of
constituents of the charmonium undergo boosts that render all of the
momenta nearly parallel. When we explicitly consider those boosts, it
follows that, for the charmonium with momentum $P_1$ ($P_2$), the plus
(minus) components are boosted by a factor of order $Q/m_c$ and the
minus (plus) components are boosted by a factor of order $m_c/Q$, while the
transverse components are unchanged. Thus, all of the momenta of the
constituents of the charmonium with momentum $P_1$ ($P_2$) are dominated
by the plus (minus) component, up to corrections of order $m_c/Q$. This
result holds provided that all of the components are of approximately
the same size in the charmonium rest frame. This is the case for $P_i$,
$q_i$, but it is also true for other momenta that characterize the
charmonium in its rest frame, such as a typical potential-gluon momentum
(which has spatial components of order $m_cv$ and a temporal component
of order $m_cv^2$) or a typical rest-frame soft-gluon momentum (which
has all components of order $m_cv$). Therefore, we use a modified soft
approximation for the constituents of the charmonium in the jet $J_i$ in
which we replace $g_{\mu\nu}$ in the gluon-propagator numerator with
$k_\mu n_{i\nu}/k\cdot n_i$, where $n_1$ ($n_2$) is a vector with unit
component in the plus (minus) direction and all other components equal
to zero. This modified soft approximation differs from the standard soft
approximation for each constituent of a quarkonium by terms that are
suppressed as $m_c/Q$. Hence, it accounts for all soft (logarithmic)
singularities, up to terms that are suppressed as $m_c/Q$. We can use
the modified soft approximation, plus the diagrammatic  Ward identities
to decouple the gluons with soft momenta from the charmonium jets,
relying on the fact that the $c$ and $\bar c$ in each charmonium are in
a color-singlet state
\cite{Collins:1981uk,Collins:1985ue,Collins:1988ig}. Once the gluons
with momenta collinear to the jet have been decoupled from $H$ and the
gluons with soft momenta have been decoupled from the charmonium jets,
there remain gluons with momenta collinear to the jet that begin and end
within the jet. In the rest frame of the charmonium, these gluons
correspond to soft and potential gluons. The contributions from regions
of momentum of order $m_cv$ or less that arise from these gluons
(including infrared divergences) can be absorbed into nonperturbative
NRQCD matrix elements \cite{Bodwin:1994jh}. The associated NRQCD
short-distance coefficients then contain only contributions involving
momenta of order $m_c$. The NRQCD short-distance coefficients couple to
the production process only through the active $c$ and $\bar c$ lines.
Thus, we have arrived at the factorized form.

As we have mentioned, the modified soft approximation allows us to
decouple gluons with soft momentum from the charmonium jets, up to
corrections that are suppressed as $m_c/Q$. Thus, in general, one 
would expect corrections to the factorized form to appear at order $m_c/Q$.
However, in the case of double-charmonium production in $e^+e^-$
annihilation, $S$ decouples from both the collinear-to-plus charmonium
jet and the collinear-to-minus charmonium jet. Each decoupling holds up
to corrections of order $m_c/Q$, and, so, the overall decoupling holds up
to corrections of order $m_c^2/s$. In perturbation theory, the
factorization-violating corrections may be enhanced by logarithms of
$s/m_c^2$. Furthermore, they are infrared divergent. In reality, these
infrared divergences are cut off by nonperturbative effects associated
with confinement. Our analysis does not determine the size of these
factorization-violating corrections: it shows only that they are
proportional to one or two powers of $m_c/Q$. The constant of
proportionality might be determined through experiment and/or lattice
calculations.

At the lowest order in $\alpha_s$ and $v$, the decoupling of soft
gluons from each charmonium is exact. At the lowest order in $\alpha_s$,
the charmonium has only the $c$ and $\bar c$ as constituents. At the
lowest order in $v$, one sets $q_i=0$, and the $c$ and $\bar c$ momenta
become equal. Then, one can apply the same soft approximation for both
the $c$ and $\bar c$ in the charmonium. Since, at the lowest order in
$\alpha_s$, there is, at most, one soft gluon, the soft decoupling in
double-charmonium production is exact if either charmonium is treated at
the lowest order in $v$.
An explicit calculation of the
one-loop corrections to $S$-wave charmonium production in $B$-meson
decays at the lowest order in $v$ \cite{Chay:2000xn} and an explicit
calculation of the one-loop corrections to $\sigma[e^+e^- \to
J/\psi+\chi_{cJ}]$, in which the $J/\psi$ is treated at lowest order in
$v$ \cite{Zhang:2008gp}, confirm our expectation that these corrections
should be free of infrared divergences. $P$-wave charmonium production in
$B$-meson decays was considered at the one-loop level in
Ref.~\cite{Song:2003yc} within the factorization framework of
Ref.~\cite{Beneke:2000ry}. When we use the light-cone distribution
amplitudes that are specified in Ref.~\cite{Beneke:2000ry}, we obtain
agreement with the calculations in Ref.~\cite{Song:2003yc}, which yield
only infrared divergences that are suppressed as $m_c^2/s$. As we have
discussed, we find more generally that the violations of factorization
are suppressed as $m_c/\sqrt{s}$. At higher orders in $v$, we expect the
soft cancellation for each charmonium to hold only up to corrections of
order $m_c/\sqrt{s}$, even in the case of $S$-wave charmonium production.
Similarly, in relative order $\alpha_s^2$ and beyond, we expect the soft
cancellation for each charmonium to hold only up to corrections of order
$m_c/\sqrt{s}$, even at the lowest order in $v$.

\begin{acknowledgments}
We thank Kuang-Ta Chao, Jianwei Qiu, George Sterman, and Yu-Jie Zhang
for helpful discussions. The work of G.T.B. and X.G.T. was supported
by the U.S. Department of Energy, 
Division of High Energy Physics, under contract DE-AC02-06CH11357.
J.L. was supported by KOSEF under contract R01-2008-000-10378-0.
\end{acknowledgments}

\end{document}